\documentclass[%
 reprint,
 amsmath,amssymb,
 aps,
floatfix,
]{revtex4-1}

\usepackage{graphicx}
\usepackage{dcolumn}
\usepackage{bm}
\usepackage{color}
\usepackage{wrapfig}
\usepackage{physics}
\usepackage{fancyhdr}
\usepackage{lastpage}
 
\newcommand{\be}{\begin{equation}}
\newcommand{\ee}[1]{\label{#1} \end{equation}}

\def\_#1{\textsubscript{#1}}
\def\^#1{\textsuperscript{#1}}

\begin{document}

\preprint{APS/123-QED}

\title{Self-stabilizing laser sails based on optical metasurfaces}

\author{Joel Siegel}
\affiliation{Department of Physics, University of Wisconsin-Madison, Madison WI 53606 USA}

\author{Anthony Wang}
\affiliation{Department of Physics, University of Wisconsin-Madison, Madison WI 53606 USA}
\altaffiliation{Department of Physics, University of Wisconsin-Madison, Madison WI 53606 USA}
\author{Sergey G. Menabde}%

\affiliation{School of Electrical Engineering, Korea Advanced Institute of Science and Technology, Daejeon 34141, Korea}

\author{Mikhail A. Kats}

\affiliation{
Department of Electrical and Computer Engineering, University of Wisconsin-Madison, Madison WI 53606 USA
}%
\author{Min Seok Jang}
\affiliation{%
School of Electrical Engineering, Korea Advanced Institute of Science and Technology, Daejeon 34141, Korea
}%

\author{Victor Watson Brar}
\thanks{vbrar@wisc.edu}%
\affiliation{%
Department of Physics, University of Wisconsin-Madison, Madison WI 53606 USA
}%

\date{\today}

\begin{abstract}

This article investigates the stability of `laser sail'-style spacecraft constructed from dielectric metasurfaces with areal densities $<$1g/m$^2$.  We show that the microscopic optical forces exerted on a metasurface by a high power laser (100 GW) can be engineered to achieve passive self-stabilization, such that it is optically trapped inside the drive beam, and self-corrects against angular and lateral perturbations.  The metasurfaces we study consist of a patchwork of beam-steering elements that reflect light at different angles and efficiencies.  These properties are varied for each element across the area of the metasurface, and we use optical force modeling tools to explore the behavior of several metasurfaces with different scattering properties as they interact with beams that have different intensity profiles.  Finally, we use full-wave numerical simulation tools to extract the actual optical forces that would be imparted on Si/SiO\_{2} metasurfaces consisting of more than 400 elements, and we compare those results to our analytical models.  We find that under first-order approximations, there are certain metasurface designs that can propel `laser-sail'-type spacecraft in a stable manner.

\end{abstract}

\maketitle

\section{\label{sec:level1}Introduction} 

\begin{figure}[h]
\includegraphics[width = 1\linewidth]{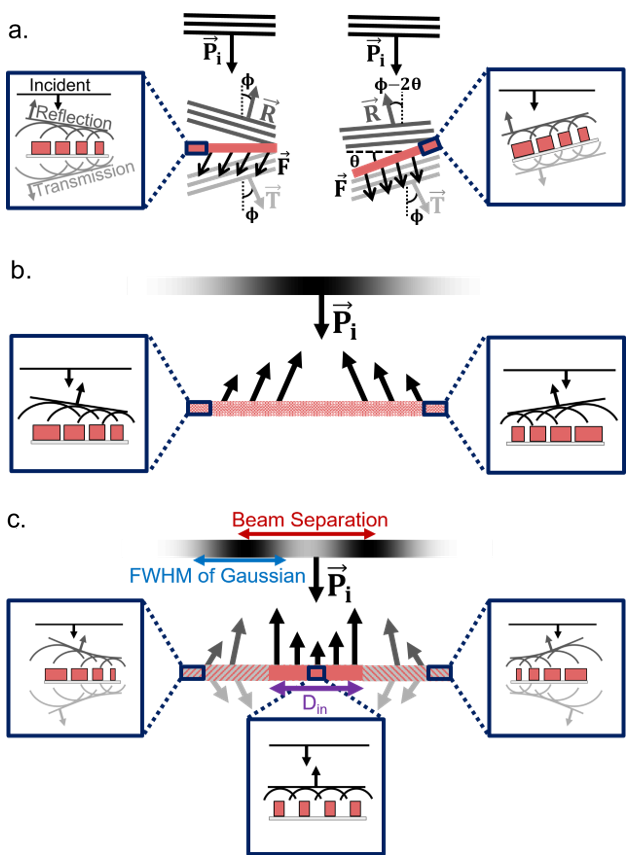}

\caption{(a) Schematic of dielectric metasurface beam-steerer that deflects an incoming beam at a constant angle, $\phi$. In dark blue, the resultant forces created by radiation pressure on a beam-steering metasurface that is flat (upper left) and tilted by an angle $\theta$ (upper right). (b) Cartoon schematic of a `V'-type sail with a Gaussian input beam. (c) Cartoon schematic of an ICE metasurface/sail with a double Gaussian input beam. 
\vspace{-8mm}}
\label{fig:force}
\label{fig:donut-sail}
\end{figure}

The optical properties of a material can be dramatically altered by structuring the material on sub-$\lambda$ length-scales to create a `metamaterial' or - for flat geometries - a `metasurface.'  In these systems, the reflection and refraction of the macroscopic light is controlled by engineering the local, microscopic scattering properties. Metasurfaces can be made much thinner than standard optical elements, and recent advances in nano-fabrication as well as optical design tools have allowed for the creation of metasurfaces that generate parabolic lenses, constant-angle beam steerers, vortex beams, and holograms,\cite{fattal2010flat,yang2014dielectric,aieta2015multiwavelength,khorasaninejad2015achromatic,zhao2016full, karimi_schulz_leon_qassim_upham_boyd_2014, cheng_ansari-oghol-beig_mosallaei_2014,decker_staude_falkner_dominguez_neshev_brener_pertsch_kivshar_2015, ma_hanham_albella_ng_lu_gong_maier_hong_2016, devlin_khorasaninejad_chen_oh_capasso_2016, lin_fan_hasman_brongersma_2014,sell_yang_doshay_yang_fan_2017, zheng_mühlenbernd_kenney_li_zentgraf_zhang_2015, ni_kildishev_shalaev_2013, huang_chen_mühlenbernd_zhang_chen_bai_tan_jin_cheah_qiu_et, west_stewart_kildishev_shalaev_shkunov_strohkendl_zakharenkov_dodds_byren_2014} with reflectivities exceeding 99$\%$ with low absorptive loss \cite{moitra2014experimental,moitra2015large}. While the beam-shaping properties of metasurfaces are well known, the optical forces present on metasurfaces have been less explored.  To understand these forces, consider Fig. \ref{fig:force}(a), which shows a standard metasurface consisting of resonators that scatter with different phases, reshaping the wavefronts of reflected and transmitted light.  Necessarily --- due to momentum conservation --- there are both normal and in-plane optical forces generated across the surface that depend on the scattering behavior. These forces are ordinarily small and inconsequential, however, as the laser power is increased, they can become large enough to impart motion on the metasurface.

One scenario where the optical forces can become large is in `laser sail' powered spacecraft, where a 100 GW beam is reflected off a mirror-like `sail', accelerating it to fraction of the speed of light.\cite{ilic_went_atwater_2018} Among the many challenges facing these efforts is the construction of the sail itself, which must display high reflectivity, minimal absorption, low weight, and be large enough such that the beam can focus on it within solar space.  A more strenuous requirement is that the sail and payload exhibit self-stability within the laser beam, such that the sail is passively steered to stay in a position of maximum thrust, and desired directionality.  In this article, we show that dielectric metasurfaces provide a promising pathway for realizing a laser sail.  We show that by locally controlling the angle and magnitude of reflection/transmission across the sail, metasurfaces can be constructed that enable both efficient propulsion, and a passive means of orientation correction that leads to self-stabilizing behavior when driven by a high power laser.

This paper is split into two parts.  We first conduct an analytical study that considers sails constructed of idealized beam-steering components and, by using dynamical modeling tools, we find deflection profiles that optimize for stability, propulsion efficiency, and operational tolerance.  Second, we use finite difference time domain (FDTD) simulation tools to model the light-scattering behavior of actual dielectric metasurfaces that are designed to match the optimized force parameters, and we compare how close realistic metasurfaces can match idealized structures.  

We note that there are several requirements of a laser sail that we do not directly address in this work.  These include potential bending and folding of the sail, the ability of the sail to be tolerant of relativistic doppler shifting, the possibility of sail overheating, and the ability of the sail to act as an antenna to transmit data back to Earth.  Those issues are discussed qualitatively at the end of this manuscript, but a comprehensive resolution is left for future studies.

\vspace{-10pt}
\section{\label{sec:level2}Motion of idealized metasurfaces} 
\vspace{-10pt}
\subsection{Dynamical Force Coefficients}
\vspace{-10pt}

In the simple metasurface beam-steering geometry shown in Figure \ref{fig:force}(a), the scattered fields from individual optical resonators pick up a linear phase gradient that results in reflected and transmitted wavefronts that are tilted at an angle $\phi$ with respect to the surface. \cite{yu_capasso_2014,yu_genevet_kats_aieta_tetienne_capasso_gaburro_2011,holloway_kuester_gordon_ohara_booth_smith_2012}. If the structure is rotated by an angle $\theta$, the optical path length of reflected light changes linearly across the surface, resulting in an additional $-2\theta$ in the angle of the reflected light. The transmitted (refracted) light, meanwhile, maintains the same angle of transmittance as the metasurface is rotated.  As the light is reflected/refracted, nonzero tangential ($F_x$) and normal ($F_z$) forces are imparted on the structure due to momentum conservation of the combined incoming/outgoing photon and metasurface system. For a generalized metasurface interacting with a non-uniform beam (i.e. Gaussian, flat top, donut, etc..), the local beam power, $P_i$, as well as $\vec{R}$ and $\vec{T}$ will all be functions of position across the metasurface, and the optical forces will have a non-trivial dependence on the metasurface's rotational or lateral offset relative to the incoming beam.  By integrating the optical force and torque contributions across the sail at different relative positions and angles of the sail and drive beam, the first order equations of motion can be derived as, 

\begin{equation}
F = m \frac{\partial^2 \delta}{\partial t^2} = C_1\delta+C_2\theta 
\label{eq:force}
\end{equation}
\begin{equation}
\tau = I \frac{\partial^2 \theta}{\partial t^2}= C_3\delta +C_4\theta
\label{eq:torque}
\end{equation}
where $F$ and $\tau$ are the lateral force and torque acting on the sail, respectively, and $\delta$ and $\theta$ are the lateral and angular offsets, respectively.  $C_{1,2,3,4}$ are the first order dynamical force coefficients ($\frac{\partial F}{\partial \delta},\frac{\partial F}{\partial \theta},\frac{\partial\tau}{\partial \delta},\frac{\partial \tau}{\partial
\theta}$) which are specific for each combination of incident beam and metasurface profile.  In this work, we considered metasurfaces that are 4m wide, with a mass, $m$, and moment of inertia, $I$, of 8.5$g$ and 17$gm^2$.  These parameters were extracted from the actual metasurfaces that are described in more detail in Section \ref{sec:level3}.  The coefficients $C_{1,2,3,4}$ were derived by calculating the total forces and torques on the metasurfaces as they are shifted over 2cm, and tilted by up to 0.1$^{\circ}$. We find that the linear approximation can be valid for offsets on the order of 10s of centimeters and rotations on the order of degrees, but there is variability between sail and beam combinations. 

The dynamical force analysis conducted in our work was restricted to a 2D model of the system, with the motion of the sail was constrained to offsets along the x and z axes and rotations about the y axis.  This approximation is made so that the sails can be faithfully modeled within our computational restraints in Section \ref{sec:level3}. We note that this 2D model does not differ significantly from a full 3D model for three reasons. First, assuming a cylindrically symmetric system in 3D, the dynamic force coefficients for the motion along y and rotation about the x axis are identical to the reversed case. Second, we do not assume the sail is spinning about the z axis, so the two types of motion are uncoupled up to a first order approximation.\cite{manchester2017stability}.  Third, while the dynamic force coefficients change when converting from a 2D model to a 3D model of the system, the general trends and approximate magnitudes of the coefficients remain the same (see Supplemental Materials), allowing for conclusions drawn in the 2D case to remain applicable to the 3D situation.

\vspace{-10pt}
\subsection{Motion Simulation} 
\label{sec:simulation}
\begin{figure}[h]
\includegraphics[width =1\linewidth]{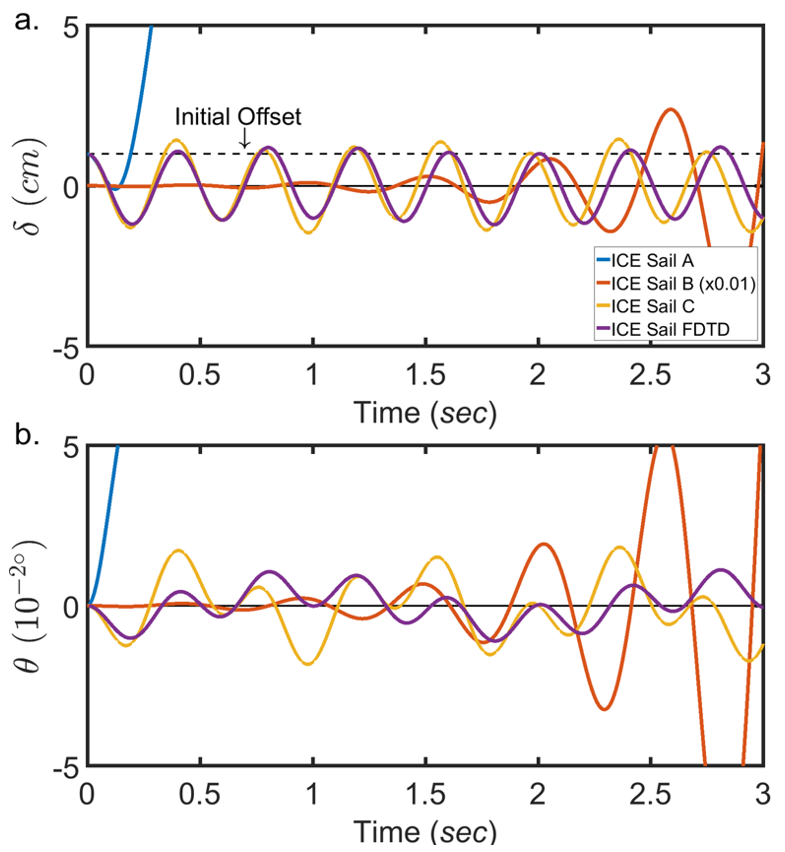}

\caption{Plots of the displacement (a) and rotation (b) for three different sail and beam combinations over a three second time interval. Each sail is four meters wide with an initial offset of 1 cm. The incident beam power is 100GW. For the three ICE sails, R$_{out}$=0.3, D$_{in}$ = 2m, and the incident beam is a double Gaussian with a 1.65 meter FWHM. ICE Sail A is for an incident beam with Beam Separation = 1.8 m, ICE Sail B is for an incident beam with Beam Separation = 2.56 m, and ICE Sail C is for an incident beam with Beam Separation = 2.40 m. ICE Sail FDTD is the motion of an FDTD simulated sail (see Section \ref{sec:level3} for details).}
\label{fig:motion}
\end{figure}

\vspace{-10pt}
Using equations \ref{eq:force} and \ref{eq:torque}, we use a `Leapfrog' integration method \cite{birdsall_langdon_1985} to simulate the motion of a metasurface in a particular beam profile in the presence of beam intensity fluctuations, or with initial lateral and angular offsets of 1$cm$ or 0.05$^{\circ}$. We find that the sail motion can display a large range of behavior depending on the metasurface structure, as well as the drive beam profile, and these behaviors can be described as either `stable' or `unstable'.   We classify a sail structure and beam combination as `stable' if, during a 60 second period, the sail does not move or rotate beyond    $2 \text{ cm}$ or $0.1^\circ$, respectively, where the linear approximation is valid.  Qualitatively, `stable' sail behavior is manifested as small oscillations about the origin, while `unstable' behavior is characterized as the sail quickly diverging in position and/or angle, as seen in Figure \ref{fig:motion}.

In order to illustrate how metasurface structure can impart stability on a laser sail, consider the sail geometry shown	in Fig. \ref{fig:force}(b).   This design --- referred to as a `V'-type sail --- reflects light at two constant angles that are equal and opposite on the left and right sides of the sail. When the sail moves to the left (right) relative to an incident Gaussian beam, higher power strikes the right (left) side of the sail, and a force is imparted that pushes the sail back to the right (left), thus establishing a lateral restoring force.   This sail geometry, however, is not stable against rotational offsets, which force the sail to be pushed out of the beam.  An advantage of a metasurface sail, however, is that it can exhibit complex reflection and transmission behaviors without distorting the geometrical shape of the surface or changing the constituent materials; as a result, beneficial force profiles can be generated in realistic structures.  As an example, in Figure \ref{fig:force}(c), we show an `inverted cat eye' (ICE) type sail that consists of a highly reflective (R$_{in}$ = 0.95) inner region that reflects the beam normal to the surface, and a more transmissive outer region (R$_{out}$) that acts as a `metalens' with a parabolic reflection/transmission profile.  When placed in a `donut' beam --- modeled as two offset Gaussians in 2D -- the ICE geometry can correct for both lateral and rotational offsets, as we show below.

Figure \ref{fig:motion} illustrates the motion of four different ICE sail and beam combinations for the same initial offset of 1 $cm$.  These behaviors can be understood by considering the dynamic force coefficients. ICE sail `A', shown in Fig. \ref{fig:motion}, has $C_1<0$, suggesting stability against lateral offsets, but it also has $C_{2,3}>0$ and $C_4 \sim 0$, indicating that: (1) positive angular rotations create positive lateral forces ($C_2>0$); (2) positive lateral offsets create positive torque ($C_3>0$); and (3) positive rotations create minimal torque ($C_4 \sim 0$). These conditions create a positive feedback effect between rotation and offset, which acts to quickly destabilize the sail.  This can be described quantitatively by considering that $C_{1}C_{4}+C_{2}C_{3}<0$ is a necessary condition for marginal stability \cite{manchester2017stability}, and for $C_4 \ll 1$ this condition cannot be satisfied when $C_{2,3}>0$.

In contrast, by changing the beam separation of the double Gaussian beam, a \textit{negative} $C_3$ value can be realized, meaning positive offsets create negative torques, and vice versa.  This provides a route towards a stable sail motion where positive rotational offsets drive positive lateral motion ($C_2 >0$) which, in turn, creates a negative torque ($C_3 <0$) that corrects for the initial rotation, leading to small oscillations about the origin (Figure \ref{fig:motion}, ICE Sail `C', yellow line).   We note, however, that while the $C_{1}C_{4}+C_{2}C_{3}<0$ condition is generally satisfied when $C_3<0$, this is not a sufficient condition to predict stable sail behavior, which depends on the relative ratios of $C_{1,2,3,4}$ within a narrow range of parameter space. ICE sail `B', for example, has a \textit{negative} $C_3$, but it displays an oscillatory motion with increasing magnitude, and does not achieve stability.

\subsection{Metasurface Stability Optimization}

\begin{figure*}
\centering

\includegraphics[width = 1\linewidth]{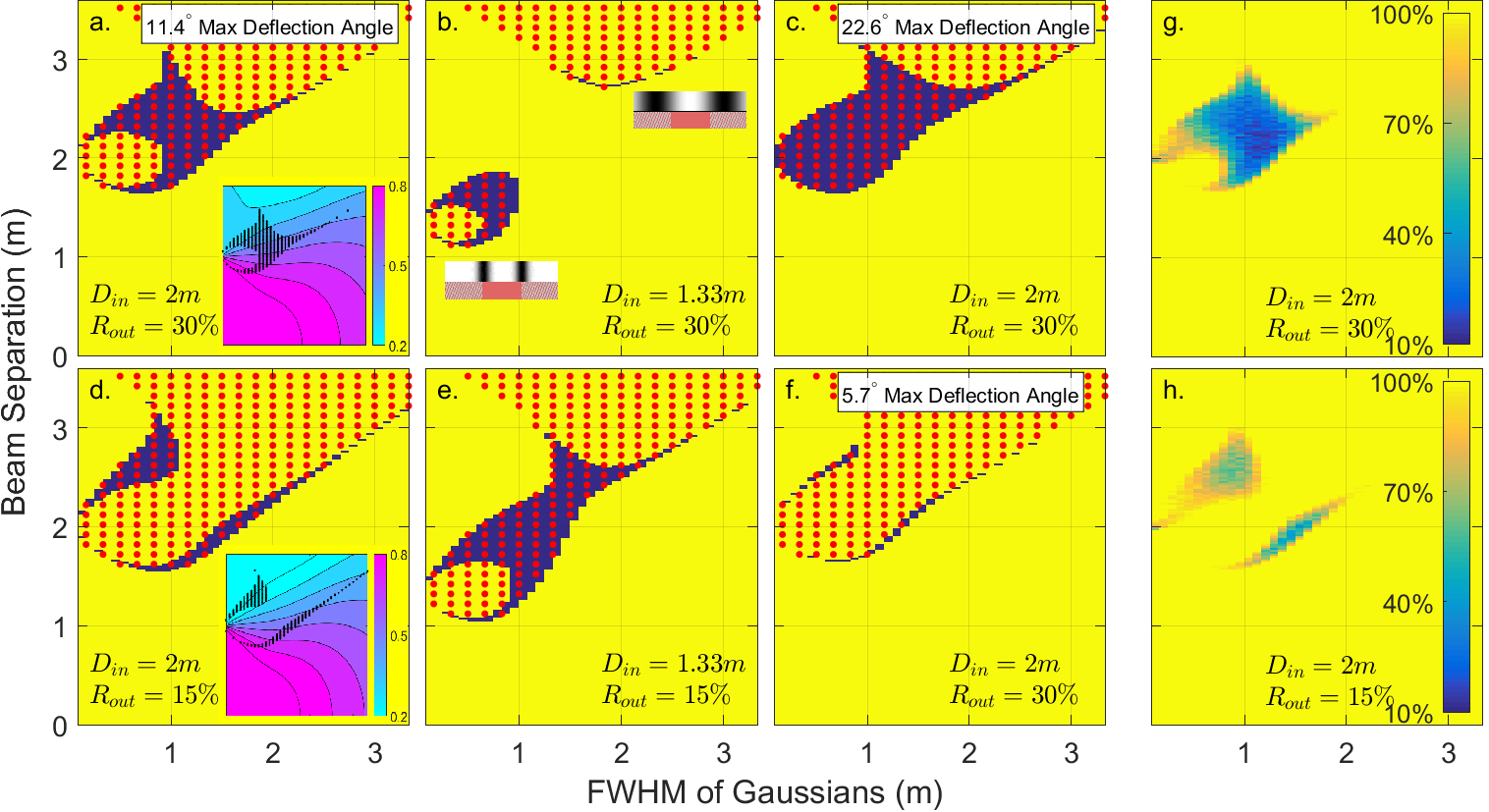}

\caption{(a-f) Analysis of stability for 6 ICE-sail configurations driven by double-Gaussian beams with varying beam separation and FWHM for an initial offset of 1cm, 50\% of the maximum allowable offset for stability. Yellow (purple) regions indicate sails configurations that are unstable (stable). The dotted red areas indicate sail configurations that satisfy the $C_{1}C_{4}+C_{2}C_{3}<0$ condition. (a,b), are ICE sails R$_{out}$ = 0.3  and D$_{in}$ = 2 m, 1.33m, respectively. Insets in (a,d) show the fraction of maximum thrust achievable for each sail/beam combination and the dotted black areas indicate sails that were stable when initially offset by 1cm. Insets in (b) show a cartoon of the beams on the sail in those regions.  (d,e) are ICE sails with R$_{out}$ = 0.15 and D$_{in}$ = 2 m, 1.33m, respectively. (c,f) are ICE sails with D$_{in}$ = 2 m and R$_{out}$ = 0.3 (similar to a), but their angle of deflection at each point is halved or or doubled in comparison to (a), respectively. The motion was simulated over a 60 second period. (g,h) Analysis of stability for two ICE-sail configurations both with D$_{in}$=2m or R$_{out}$=30$\%$ (g) and R$_{out}$=15$\%$ (h) reflection efficiencies for their outer regions. The motion of the sail is tracked over 5 minutes with 0.12\% noise introduced to the beam.  Yellow regions indicate sails that failed 100\% of the time.  Blue indicates regions that maintained the lowest rates of failure.} 

\label{fig:stability_plots}
\end{figure*}

In order to search for and classify stable ICE sail configurations, we simulated the motion of 4 meter wide sails with varying reflection coefficients for the outer region (R$_{out}$ = $0.15, 0.3$), inner region widths (D$_{in}$ = $1.33 \text{ }m,\text{ } 2\text{ } m$), and outer edge deflection angle ($\theta_{max} =   5.7^{\circ}, 11.4^{\circ} \text{ and } 22.6^{\circ}$) (an expanded analysis is shown in the Supplementary Materials).  We model the motion of these sails in a continuum of double-Gaussian beam profiles with an incident beam composed of two Gaussian beams symmetrically offset from the center, with the beam defined by the FWHM of the Gaussians, and the separation between them.  The results of these simulations are shown in Fig. \ref{fig:stability_plots}.

These results demonstrate that stability is strongly dependent on the metasurface profile, as well as the beam shape. Overall, as R$_{out}$ decreases and D$_{in}$ increases, the number of regions that satisfy $C_{1}C_{4}+C_{2}C_{3}<0$ increases, a result that occurs because these sails can achieve negative values of $C_3$ for more beam combinations. The number of configurations that show actually stable sail behavior, however, does not show a strong, intuitive dependence on R$_{out}$ or D$_{in}$.  There is, however, a stronger dependence on the outer edge deflection angle.  Figures \ref{fig:stability_plots} (a), (c), and (f) compare the stability behaviors of a sails that have identical $R_{out}$ and D$_{in}$ values, but with different outer edge deflection angles.  These figures indicate that steeper deflection angles (or, equivalently, parabolic lens profiles with shorter focusing distances) allow for more stability conditions, and loosen constraints on the drive laser.

In order to determine which sail/beam configurations are `maximally stable', it is necessary to consider beam intensity profiles that contain time dependent distortions which, in real world scenarios, could be caused by laser interference (`speckle') and atmospheric fluctuations, which can destabilize the sail.   Here we analyze the sensitivity of metasurface sails to such perturbations by using a Monte Carlo method to introduce randomized, time-varying intensity fluctuations in the beam profile while simulating the motion of the sail.  We assume intensity variations that occur over 10 cm characteristic lengthscales, with time correlations of 1ms, corresponding to timescales associated with atmospheric turbulence \cite{deserno}. With the fluctuation intensity set to 0.12\%, we ran 100 simulations for each configuration show in Fig.\ref{fig:stability_plots} (a) and (d) and we recorded the probability that a sail maintained stability.  The result of those simulations are shown in Figure \ref{fig:stability_plots} (g) and (h), where we observe a large variation in stability rates between sail configurations. When R$_{out}$ =0.3 (g), the stable region forms a large basin of stability that achieve a failure rate of as low as 13\%. If we decrease R$_{out}$ to 0.15 (h), we increase our minimum failure rate to 40\%. 

These simulations also provide insight for deducing which stable configuration provide maximum thrust, which is plotted in the insets of Fig. \ref{fig:stability_plots} (a) and (d).  Considering only beams that have a FWHM $>$ 1.6m (i.e. realistic, diffraction-limited beams) , the maximum thrust achievable with a stable sail/beam configuration is 370 N and 400 N for $R_{out} = 0.3$ and $0.15$, respectively.  For comparison, the maximum theoretical thrust -- which occurs when 100\% of the light is reflected normally, is 667 N.

\section{\label{sec:level3} Full-Wave Simulations}

In Section \ref{sec:level2} we modeled the behavior of theoretical beam-steering metasurfaces with idealized optical properties.  Actual dielectric metasurfaces, however, rely on using dielectric nanostructure arrays designed to control the optical wavefronts of the scattered laser fields, and these structures diverge from ideal behavior due to phase slips in the structure, phase-dependent reflectivity/transmissivity, and interactions between nano-resonators.  These effects lead to sub-optimal efficiencies, and can potentially create large modifications in the dynamical force coefficients of a metasurface laser sail.  In this section, we design and model large-scale ICE metasurfaces and we extract the optical forces on those structures as they are tilted and offset in a drive beam.  We then compare the dynamical force coefficients of an actual dielectric surface to those of an idealized structure, and show that the self-stability behavior described in Section \ref{sec:level2} is achievable in real-world structures.

\begin{figure}[ht]
\includegraphics[width = 1\linewidth]{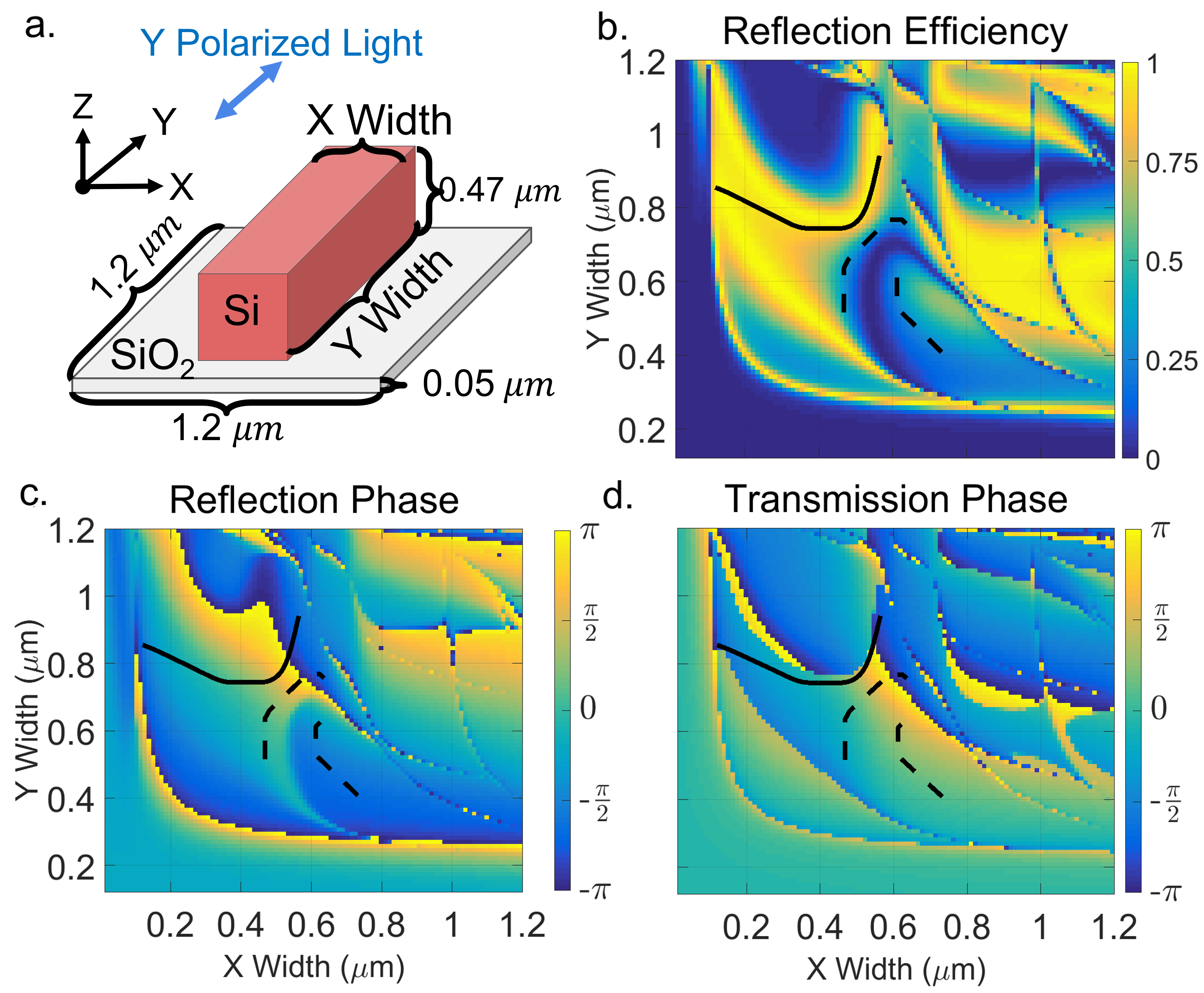}

\caption{Diagram of the unit cell, reflected magnitude/phase, and transmission phase. (a) Metasurface unit cell consisting of a rectangular block of Si on a SiO\_{2} surface. The boundary conditions in the X and Y direction are periodic. (b) Magnitude of reflected light as the X and Y dimensions of the Si block vary. (c) Phase of reflected light as the X and Y dimensions of the Si block vary. (d) Phase of transmitted light as the X and Y dimensions of the Si block vary. The solid (dashed) lines indicate a 95\% (30\%) reflectivity path covering 2$\pi$ phase}.

\label{fig:unit_cell}
\end{figure}

\subsection{Optical Design}

\begin{figure}[t]
\centering
\includegraphics[width = 1\linewidth]{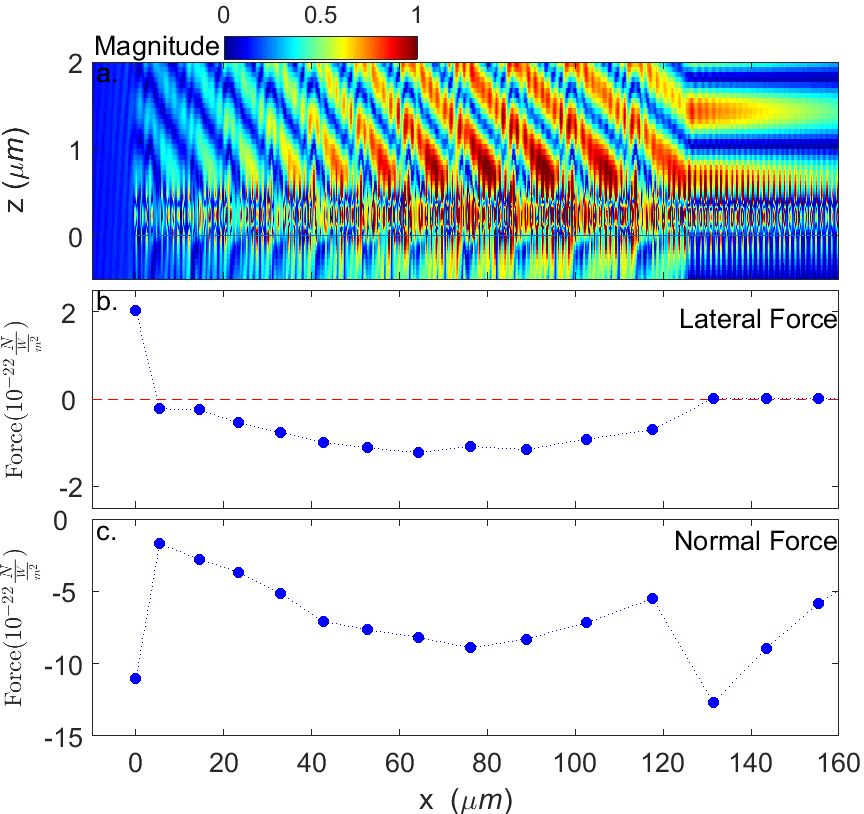}
\caption{(a) Electric Field Profile of half of a ICE sail with a reflective (95\% reflectivity) inner radius of 127 $\mu m$ and an outer transmissive (30\% reflectivity) region with a parabolic scattering profile where the inner edge steers at 2.8$^\circ$ and the outer edge steers at 5.7$^\circ$. (b,c) The local lateral and normal components of the optical forces on the sail, calculated by integrating the Maxwell stress tensor (MST) over a surface that encloses local groups of resonators.  For the outer region the bounding surfaces contain resonator groups that are between phase slips which ranges from five resonators for the outer edge, 10 resonators inner edge. For the inner, reflective region, the bounding surfaces contains 10 resonators each.  The force is normalized using an input beam power of 1 Watt across the sail. }
\label{fig:FieldProfileDoughnut5}
\end{figure}

\begin{figure}

\includegraphics[width=1\linewidth]{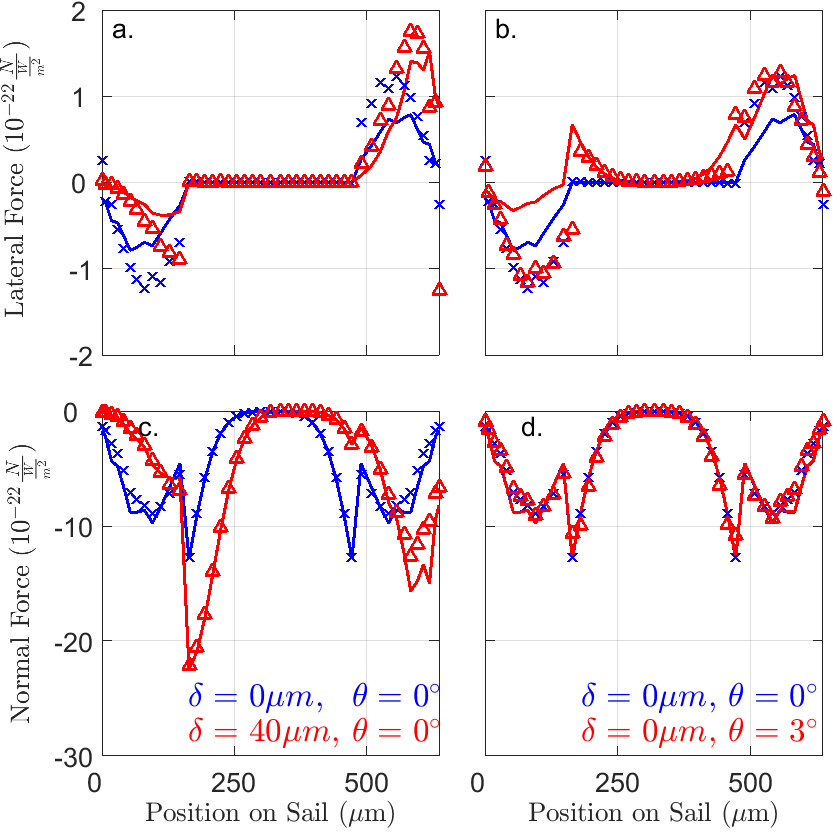}
\caption{The local lateral (x) and normal (z) components of the optical forces on the entire sail shown in Figure \ref{fig:FieldProfileDoughnut5}, calculated from the Full-Wave simulations (blue xs and red triangles) or calculated using the methods described in Section \ref{sec:level2} (solid blue/red lines). (a) corresponds to the force in the x direction when the sail is offset. (b) corresponds to the force in the x direction when the sail is rotated. (c) correspond to the force in the z direction when the sail is offset. (d) corresponds to the force in the z direction when the sail is rotated. In all plots, the un-offset, unshifted force is shown for comparison in blue and the offset or rotated force is shown in red.}
\label{fig:comparison}
\end{figure}

In order to achieve full control of optical wavefronts, it is necessary to engineer a set of dielectric resonators that can scatter light with phase shifts ranging from $0$ to $2\pi$, and with arbitrarily small or large reflection/transmission efficiencies.  Here we use Si nanoresonators on SiO$_2$, which have previously been shown to be effective in the construction of metasurfaces with high reflectivity  \cite{moitra2014experimental,moitra2015large}  as well as metasurfaces that act as efficient focusing optics.  In order to minimize loss, we assume a drive beam wavelength of 1.55 $\mu m$.  The set of resonators we use in this work are shown in Figure \ref{fig:unit_cell}(a); 470-nm-high Si blocks with variable length and width are placed on a 50-nm-thick SiO$_{2}$ substrate, and the spacing between resonators (center to center) is maintained at a constant 1.2 $\mu m$.  The scattering properties of these resonators (and all future metasurfaces described in this work) are calculated using commercial finite-difference time domain tools (Lumerical FDTD).  As the resonator dimensions are varied, the magnitude and phase of the reflected/transmitted fields also change, as shown in Figure \ref{fig:unit_cell}(b-d).  Contours of constant reflectivity that cover 2$\pi$ phase can be selected from these plots.  Two examples are shown in Figure \ref{fig:unit_cell} as solid and dashed lines for 95\% and 30\% reflectivities, respectively.  Note that for 95\% reflectivity a single continuous path can be chosen, while for 30\% reflectivity, two separate paths are required for full phase coverage.  Similar contour selection can be performed for reflectivities ranging from 95-15\%.

The sets of resonators described above can be used to construct metasurfaces as outlined in Sections \ref{sec:level1} and \ref{sec:level2}, where the magnitude and direction of the reflected/transmitted wavefront is locally controlled by placing resonators with the desired scattering profiles at each location on the surface.  This is demonstrated in Figure \ref{fig:FieldProfileDoughnut5} where we have plotted the electric field profile of the scattered waves from one side of an  ICE sail constructed using the nanoresonators in Fig. \ref{fig:unit_cell}.  For this metasurface and all others studied below, periodicity is assumed along the y-dimension.  The structure has a total width is 504 $\mu m$ containing 420 individual resonators with constant period of 1.2 $\mu m$.  The input beam is a `double Gaussian' beam, with a FWHM of 82.9 $\mu m$ and an annular diameter of 334 $\mu m$. The resonators placed from 0 to 125 microns were chosen from the dashed black paths in Figure \ref{fig:unit_cell} and provide 30\% reflection, along with a parabolic beam-steering profile that ranges from 5.7$^\circ$ at the edge of the sail to  2.8$^\circ$ at the boundary between the outer and inner regions. The inner region is formed by 95\% reflective resonators that all have the same phase, which is chosen to match the phase of the innermost resonators of the outer 30\% reflective region. Details of how the resonators are chosen at each position are provided in the Supplemental Materials. 

\subsection{Optical Forces}

In order to exhibit the self-stabilized behavior described in Section \ref{sec:level2}, it is necessary to design metasurface sails that faithfully generate particular dynamical force coefficients, and that are free from perturbations that could lead to localized folding.  Here, we calculate the optical forces locally by by integrating the Maxwell Stress Tensor (MST) around boxes enclosing individual resonators or small groups of resonators.  We then sum the vector components of those local forces to calculate the overall lateral and normal forces, as well as the torque on the sail as it is tilted and displaced within the beam.  In Fig. \ref{fig:FieldProfileDoughnut5} (b,c) we plot an example of these force components for a metasurface ICE sail illuminated by a double Gaussian beam.  These forces display the following expected general trends: (1) there is a lateral force on the outer region that stretches the sail and scales with steering angle and beam intensity; (2) there is no lateral force on the center region, which is designed to be strictly normally reflecting; (3) there is a normal force across the sail that scales with beam intensity and sail reflectivity.  In this example, we have integrated the MST over a bounding box that groups resonators between phase slips in the metasurface, such that each x,z force shown in Fig. \ref{fig:FieldProfileDoughnut5}(b,c) is determined by calculating the net optical force on groups of 5-10 resonators, depending on the local steering angle.  This is done to decrease calculation time, and it has a negligible effect on the net torque or force on the sail, which was confirmed by comparing to calculations where the MST was integrating over individual resonators.  

When forces are analyzed on individual resonators, it is observed that there are discontinuities in the force profile that occur due to phase slips and interactions between pairs of resonators, which can cause to actual scattering phase to diverge from the predicted phase. Those effects --- which cause anomalies in the scattered E-field profile visible in Fig. \ref{fig:FieldProfileDoughnut5}(a) --- are discussed more in the Supplementary Materials.  In most cases those local discontinuities create forces that are several orders weaker than the elastic restoring forces in the underlying SiO$_2$ slab.  However, even when averaged over several resonators, the effects of phase shifts due to resonator interactions can be observable.  For example, the depression in lateral forces for the resonators located at $\sim$78$\mu m$ in Fig. \ref{fig:FieldProfileDoughnut5}(b) is due to such  interactions.  Moreover, the lateral force reverses sign at the end of the sail due to diffraction from the sail edge, and from the altered scattering properties of the last resonator, which is in an asymmetric environment.  For small steering angles, these effects have a minor contribution to the overall forces on the sail, however, as beam steering angles are increased so is the frequency of phase slips, which leads to larger contributions to the overall lateral force.  Methods for potentially ameliorating and/or accommodating for such effects are discussed in the `Conclusions' section.

In order to extract the dynamical force coefficient from our simulations, it is necessary to calculate the local optical forces as the metaurface sail is shifted and tilted within the beam, and then perform linear fits to the position/tilt vs. force/torque dependencies.  Figure \ref{fig:comparison} shows the numerically simulated local forces for offsets of 0 $\mu m$ and 40 $\mu m$ and, separately, rotations of $0^\circ$ and $3^\circ$, for the ICE sail shown in Fig. \ref{fig:FieldProfileDoughnut5}.  In these plots, we include analytical calculations (lines) of local optical forces for a sail made of ideal beamsteering components, which show good agreement with the forces extracted from the FDTD simulated metasurfaces.  Shifted simulations were performed over lateral and rotational steps of 5-10 $\mu$m and 0.5-1$^{\circ}$, respectively, and we found that the force/torque showed a linear dependence on offset/rotation angle over ranges of 20-40 $\mu$m and 1-3$^{\circ}$.  First order fits were used to determine the effective dynamical force coefficients, $C_{1,2,3,4}$, which are shown in \ref{tab:constants}.    We find that the coefficients from an actual dielectric metasurface are of the same sign and order-of-magnitude as those calculated analytically, but can vary by as much as 40\%.  These differences can be attributed to the aforementioned phase slips and inter-resonator interactions which add distortions to the reflected/transmitted phase fronts of the scattered light, and can also lead changes in the overall reflection coefficients.

\begin{table}[h]

\begin{tabular}{|p{1.6cm}||p{1.55cm}|p{1.55cm}|p{1.55cm}|p{1.55cm}|}
\hline
ICE Sail  & $C_1$ ($\frac{N}{Wm}$) &  $C_2$ ($\frac{N}{Wdeg}$) & $C_3$ ($\frac{N}{W}$) & $C_4$ ($\frac{Nm}{Wdeg}$) \\
 \hline
 \hline
 Ideal & -1.43E-6 & 6.06E-11 & -2.23E-10 & 1.03E-15 \\
 \hline
 Full-wave  & -1.79E-6 & 4.13E-11 & -2.23E-10 & 3.00E-15\\
 \hline
 Full-wave Scaled & -2.29E-10 & 4.13E-11 & -2.23E-10 & 2.36E-11\\
 \hline
\end{tabular}

\caption{Dynamic Force Coefficients of an ICE Sail. The Full-Wave dynamic force coefficients scaled up to a 4 meter wide sail are shown as well. }
\label{tab:constants}
\end{table}

\begin{figure}[h]
\includegraphics[width = 1\linewidth]{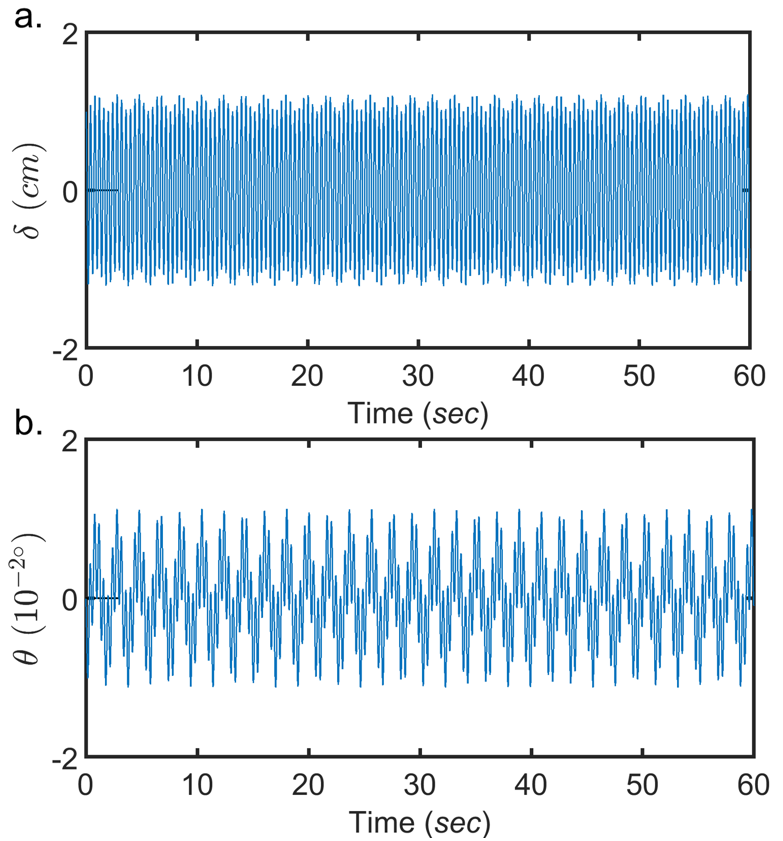}
\caption{Motion (a) and rotation (b) of an FDTD simulated sail with an initial offset of 1cm for 60 seconds.}
\label{fig:motion_lumerical}
\end{figure}

The overall stability of the ICE metasurface sail can be tested by using the dynamical force coefficients in Table \ref{tab:constants}, along with the methods described in Section \ref{sec:level2}.  However, in order to test the viability of a 4 meter metasurface, these coefficients must be scaled accordingly as the beam size (annulus and FWHM of individual Gaussians) and sail size are increased by $\sim$8000$\times$.  Such scaling has no effect on $C_2$ or $C_3$, but $C_1$ will scale inversely with size, while $C_4$ will scale linearly with size.  This behavior was confirmed by calculating $C_{1,2,3,4}$ analytically for 504$\mu$m and 4m ICE sails with equivalent ratios for reflective/transmissive regions, and equivalent steering angles.  An analysis of these scaling laws is provided in the Supplementary Materials.  The resulting scaled dynamical coefficients for a 4 meter ICE metasurface sail are given in the bottom row of Table \ref{tab:constants} and, using these values, we can model the motion of a sail with an initial lateral offset of 0.01  cm, which is shown in Figure \ref{fig:motion_lumerical}.  These results show that a metasurface sail constructed from Si nanoresonators can exhibit self-stability within a drive beam for over 300 seconds, without ever leaving the range where the linear approximation is valid and exhibiting behavior that closely resembles the motion of the idealized structures discussed in Section \ref{sec:level2}.

\section{\label{sec:conclusion} Discussion}

The combined results of Sections \ref{sec:level2} and \ref{sec:level3} show that dielectric metasurfaces offer a viable pathway for creating laser sails that maintain directionality and stability within high power laser beams.  In order to comprehend the real world speed that such a metasurface sail may achieve, we can consider a 100GW drive laser constructed from a 100km diameter ground-based array of projectors.  Operating at wavelength of 1.55 $\mu$m and for distances up of up to 10\^8 km, such a beam can maintain features with a FWHM of 1.67m, which --- in an assumed annular geometry --- would generate a thrust of $\sim$360N and allow for self-stable behavior for the sail modeled in Figure \ref{fig:motion_lumerical}.  The weight of such a sail is $\sim$8.5g, and if we assume a 5g payload, these conditions would allow for final velocities of 0.21c to be achieved in $\sim$50 minutes, after which the craft travels too far to project a laser shape that yields stability\cite{mcinnes_2004}.  In addition to self-stability and a high achievable velocity, we note that the ICE sails discussed in this work contains an outer region that can be utilized as a parabolic collimator with a 4 meter aperture, that could potentially be used to transmit a signal. 

While these results are promising, the analysis presented in this work makes several assumptions and simplifications, and ultimately the design of a metasurface laser sail will require new breakthroughs in metasurface architectures, as well more powerful simulation tools.  The key issues we would like to address are as follows:

\textbf{Folding, strain, and local forces -}  Throughout this work, we have assumed that the metasurface sail maintains a rigid flat shape, regardless of local force gradients.  In a real world situation, however, a 50nm thick SiO$_2$ substrate will easily bend and fold over both macroscopic and microscopic lengthscales.  Those perturbations will alter the optical performance of the metasurface, and second-generation designs must incorporate mitigation strategies for such effects.  The primary cause of microscopic strain --- which occurs between 2-10 nanoresonators --- is interactions between resonators, which alters their scattering properties such that the beamfronts and optical forces are distorted.  Those effects can be compensated for by using inverse design and optimization methods that consider such interactions and adjust resonator shape accordingly.  Such methods should not only allow for the creation of metasurfaces with smoother force profiles, but they could also enable higher scattering angles, which increases the lateral forces.  More problematic is the macroscopic strain, which occurs due to non-uniform beam intensity and changes in the metasurface reflectivity.  Figure \ref{fig:FieldProfileDoughnut5} (c), for example, shows that the normal forces on the sail is not constant, which will lead to bowing of the sail.  In order to correct for those effects, it is necessary to design a sail with a favorable relationships between structure and light scattering such that, for example, outwardly bowed surfaces reflect light less.  Alternatively, metasurface sails that exhibit stability within flat beam profiles would suffer less from such effects, or metasurface elements that pull outwardly on the sail can be incorporated.  We note that the ICE sail design contains some elements of this latter concept, with the edges of the sail scattering at steeper angles and generating larger lateral forces.  Finally, we note that abnormal optical forces are easily generated on the rim of the sail, where edge diffraction can drive large distortion in the scattered beam fronts.   Those effects must be compensated for by designing metasurface edges that minimize diffraction, or by using drive beams that have negligible intensity on the edge of the sail.

\textbf{Doppler shift -}  The metasurfaces discussed in Section \ref{sec:level3} are designed to work at a single fixed wavelength (1.55$\mu$m), while relativistic light propulsions requires a reflectors that works across a range of Doppler shifted frequencies.  Achieving velocities up to 0.21$c$, for example, requires a metasurface that exhibits effective propulsion and self-stability from 1.55 - 1.99 $\mu$m.  Typically, the scattering phase of an individual nanoresonator is dependent on wavelength and that dependence can lead to distorted beamfronts, and local optical forces that disturb self-stabilizing motion.  However, we note that the design of broadband, achromatic metasurfaces that shape light equally across many wavelengths is an active area of research with several recent successes \cite{wang, wang_dong_li,khorasaninejad_aieta_kanhaiya_kats_genevet_rousso_capasso_2015}, and those methods can be equally applied to the metasurface laser sail problem.  

\textbf{Residual motion - }  The force models used in Sections \ref{sec:level2} were first order, and did not include damping terms, which lead to oscillatory sail motion that did not decay in magnitude.  In order to achieve true stability, however, it is necessary to discover conditions where the amplitude of oscillation is reduced to zero, which prevents lateral motion or rotations that would persist after the drive laser is turned off, and would misdirect the spacecraft.  Optical damping terms could potentially be included through Doppler effects, however, such methods would require relativistic lateral motion, and unrealistic fabrication parameters.  More promising is to incorporate some hysteretic motion in the sail whereby, for example, some bending in the sail leads to changes in the optical scattering that is dependent on the velocity.  Such stabilization methods are used in other mechanical systems \cite{semenov_shevlyakova_meleshenko_2013}, and they are an attractive option for laser sail applications, where no obvious damping method exists. 

In conclusion, we have studied the use of dielectric metasurfaces to act as laser sails for relativistic, interstellar spacecraft.  We have shown that the ability of metasurfaces to reshape optical beamfronts can also be used to control local optical forces, and for extremely high power lasers those forces can impart an optical trapping-like effect on the metasurface.  That effect is potentially useful for stabilizing laser sails and we find a large parameter space that such behavior is expected to occur.  These results represent a new area of metasurface research that will benefit greatly from recent advances in metasurface design, and also introduces new challenges for the nanophotonics community.  

\section{Acknowledgements}
This research was performed using the compute resources and assistance of the UW-Madison Center For High Throughput Computing (CHTC) in the Department of Computer Sciences. The CHTC is supported by UW-Madison, the Advanced Computing Initiative, the Wisconsin Alumni Research Foundation, the Wisconsin Institutes for Discovery, and the National Science Foundation, and is an active member of the Open Science Grid, which is supported by the National Science Foundation and the U.S. Department of Energy's Office of Science.

S. G. M. and M. S. J. acknowledge support from Creative Materials Discovery Program through the National Research Foundation of Korea funded by the Ministry of Science and ICT (2016M3D1A1900038).

M. A. K. is supported by the Air Force Office of Scientific Research (FA9550-18-1-0146)

\bibliography{starref}

\end{document}